\documentclass[12pt]{iopart}

\usepackage{iopams}  
\usepackage{graphicx}
\usepackage{hyperref}
\usepackage{lineno}
\usepackage{color}
\usepackage{xcolor}
\usepackage{url}
\usepackage{amsmath}
\newcommand{\vect}[1]{\mbox{\boldmath${#1}$}}

\newcommand{\beq}{\begin{equation}}
\newcommand{\eeq}{\end{equation}}
\newcommand{\beqa}{\begin{eqnarray}}
\newcommand{\eeqa}{\end{eqnarray}}

\newcommand{\lmk}{\left(}
\newcommand{\rmk}{\right)}
\newcommand{\lkk}{\left[}
\newcommand{\rkk}{\right]}

\newcommand{\la}{\langle}
\newcommand{\ra}{\rangle}

\newcommand{\vs}{{\vect s}}
\newcommand{\vx}{{\vect x}}
\newcommand{\vy}{{\vect y}}

\newcommand{\vts}{\tilde{\vect s}}
\newcommand{\vtx}{\tilde{\vect x}}
\newcommand{\vty}{\tilde{\vect y}}

\newcommand{\ta}{{\tilde a}}
\newcommand{\tw}{{\tilde w}}
\newcommand{\tx}{{\tilde x}}
\newcommand{\ty}{{\tilde y}}

\newcommand{\ts}{{\tilde s}}

\newcommand{\bm}[1]{{\mbox{\boldmath $#1$}}}

\begin{document}

\begin{flushright}
RESCEU-5/22
\end{flushright}

\title{Noise subtraction from KAGRA O3GK data using Independent Component Analysis}
\author{
H.~Abe$^{1}$, 
T.~Akutsu$^{2, 3}$, 
M.~Ando$^{4, 5}$, 
A.~Araya$^{6}$, 
N.~Aritomi$^{4}$, 
H.~Asada$^{7}$, 
Y.~Aso$^{8, 9}$, 
S.~Bae$^{10}$, 
Y.~Bae$^{11}$, 
R.~Bajpai$^{12}$
% ...
\\
(KAGRA Collaboration)
}


\address{${}^{1}$ Graduate School of Science, Tokyo Institute of Technology, Meguro-ku, Tokyo 152-8551, Japan}
\address{${}^{2}$ Gravitational Wave Science Project, National Astronomical Observatory of Japan (NAOJ), Mitaka City, Tokyo 181-8588, Japan}
\address{${}^{3}$ Advanced Technology Center, National Astronomical Observatory of Japan (NAOJ), Mitaka City, Tokyo 181-8588, Japan}
\address{${}^{4}$ Department of Physics, The University of Tokyo, Bunkyo-ku, Tokyo 113-0033, Japan}
\address{${}^{5}$ Research Center for the Early Universe (RESCEU), The University of Tokyo, Bunkyo-ku, Tokyo 113-0033, Japan}
\address{${}^{6}$ Earthquake Research Institute, The University of Tokyo, Bunkyo-ku, Tokyo 113-0032, Japan}
\address{${}^{7}$ Department of Mathematics and Physics, 
Gravitational Wave Science Project, Hirosaki University, Hirosaki City, Aomori 036-8561, Japan}
\address{${}^{8}$ Kamioka Branch, National Astronomical Observatory of Japan (NAOJ), Kamioka-cho, Hida City, Gifu 506-1205, Japan}
\address{${}^{9}$ The Graduate University for Advanced Studies (SOKENDAI), Mitaka City, Tokyo 181-8588, Japan}
\address{${}^{10}$ Korea Institute of Science and Technology Information (KISTI), Yuseong-gu, Daejeon 34141, Korea}
\address{${}^{11}$ National Institute for Mathematical Sciences, Yuseong-gu, Daejeon 34047, Korea}
\address{${}^{12}$ School of High Energy Accelerator Science, The Graduate University for Advanced Studies (SOKENDAI), Tsukuba City, Ibaraki 305-0801, Japan}
\address{${}^{13}$ Department of Astronomy, Beijing Normal University, Beijing 100875, China}
\address{${}^{14}$ Department of Applied Physics, Fukuoka University, Jonan, Fukuoka City, Fukuoka 814-0180, Japan}
\address{${}^{15}$ Department of Physics, Tamkang University, Danshui Dist., New Taipei City 25137, Taiwan}
\address{${}^{16}$ Department of Physics and Institute of Astronomy, National Tsing Hua University, Hsinchu 30013, Taiwan}
\address{${}^{17}$ Department of Physics, Center for High Energy and High Field Physics, National Central University, Zhongli District, Taoyuan City 32001, Taiwan}
\address{${}^{18}$ Department of Physics, National Tsing Hua University, Hsinchu 30013, Taiwan}
\address{${}^{19}$ Institute of Physics, Academia Sinica, Nankang, Taipei 11529, Taiwan}
\address{${}^{20}$ Univ. Grenoble Alpes, Laboratoire d'Annecy de Physique des Particules (LAPP), Universit\'e Savoie Mont Blanc, CNRS/IN2P3, F-74941 Annecy, France}
\address{${}^{21}$ Department of Astronomy, The University of Tokyo, Mitaka City, Tokyo 181-8588, Japan}
\address{${}^{22}$ Faculty of Engineering, Niigata University, Nishi-ku, Niigata City, Niigata 950-2181, Japan}
\address{${}^{23}$ Department of Physics, Graduate School of Science, Osaka City University, Sumiyoshi-ku, Osaka City, Osaka 558-8585, Japan}
\address{${}^{24}$ State Key Laboratory of Magnetic Resonance and Atomic and Molecular Physics, Innovation Academy for Precision Measurement Science and Technology (APM), Chinese Academy of Sciences, Xiao Hong Shan, Wuhan 430071, China}
\address{${}^{25}$ Department of Physics, Ulsan National Institute of Science and Technology (UNIST), Ulju-gun, Ulsan 44919, Korea}
\address{${}^{26}$ Shanghai Astronomical Observatory, Chinese Academy of Sciences, Shanghai 200030, China}
\address{${}^{27}$ Institute for Cosmic Ray Research (ICRR), KAGRA Observatory, The University of Tokyo, Kashiwa City, Chiba 277-8582, Japan}
\address{${}^{28}$ Faculty of Science, University of Toyama, Toyama City, Toyama 930-8555, Japan}
\address{${}^{29}$ Institute for Cosmic Ray Research (ICRR), KAGRA Observatory, The University of Tokyo, Kamioka-cho, Hida City, Gifu 506-1205, Japan}
\address{${}^{30}$ College of Industrial Technology, Nihon University, Narashino City, Chiba 275-8575, Japan}
\address{${}^{31}$ Institute of Astronomy, National Tsing Hua University, Hsinchu 30013, Taiwan}
\address{${}^{32}$ Department of Astronomy \& Space Science, Chungnam National University, Yuseong-gu, Daejeon 34134, Korea, Korea}
\address{${}^{33}$ Department of Physical Sciences, Aoyama Gakuin University, Sagamihara City, Kanagawa  252-5258, Japan}
\address{${}^{34}$ Kavli Institute for Astronomy and Astrophysics, Peking University, Haidian District, Beijing 100871, China}
\address{${}^{35}$ Graduate School of Science and Engineering, University of Toyama, Toyama City, Toyama 930-8555, Japan}
\address{${}^{36}$ Nambu Yoichiro Institute of Theoretical and Experimental Physics (NITEP), Osaka City University, Sumiyoshi-ku, Osaka City, Osaka 558-8585, Japan}
\address{${}^{37}$ Department of Physics, Ewha Womans University, Seodaemun-gu, Seoul 03760, Korea}
\address{${}^{38}$ National Astronomical Observatories, Chinese Academic of Sciences, Chaoyang District, Beijing, China}
\address{${}^{39}$ School of Astronomy and Space Science, University of Chinese Academy of Sciences, Chaoyang District, Beijing, China}
\address{${}^{40}$ Institute for Cosmic Ray Research (ICRR), The University of Tokyo, Kashiwa City, Chiba 277-8582, Japan}
\address{${}^{41}$ Faculty of Science, University of Toyama, Toyama City, Toyama 930-8555, Japan}
\address{${}^{42}$ Department of Physics, Myongji University, Yongin 17058, Korea}
\address{${}^{43}$ Department of Computer Simulation, Inje University, Gimhae, Gyeongsangnam-do 50834, Korea}
\address{${}^{44}$ Institute of Particle and Nuclear Studies (IPNS), High Energy Accelerator Research Organization (KEK), Tsukuba City, Ibaraki 305-0801, Japan}
\address{${}^{45}$ School of Physics and Astronomy, Cardiff University, Cardiff, CF24 3AA, UK}
\address{${}^{46}$ Instituto de Fisica Teorica, 28049 Madrid, Spain}
\address{${}^{47}$ Department of Physics, Nagoya University, Chikusa-ku, Nagoya, Aichi 464-8602, Japan}
\address{${}^{48}$ Department of Physics, National Cheng Kung University, Tainan City 701, Taiwan}
\address{${}^{49}$ School of Physics and Technology, Wuhan University, Wuhan, Hubei, 430072, China}
\address{${}^{50}$ National Center for High-performance computing, National Applied Research Laboratories, Hsinchu Science Park, Hsinchu City 30076, Taiwan}
\address{${}^{51}$ Department of Physics, National Taiwan Normal University, sec. 4, Taipei 116, Taiwan}
\address{${}^{52}$ Istituto Nazionale di Fisica Nucleare (INFN), Universita di Roma "La Sapienza", 00185 Roma, Italy}
\address{${}^{53}$ Institute for Photon Science and Technology, The University of Tokyo, Bunkyo-ku, Tokyo 113-8656, Japan}
\address{${}^{54}$ Department of Physics, University of Wisconsin-Milwaukee, Milwaukee, WI 53201, USA}
\address{${}^{55}$ Institute of Space and Astronautical Science (JAXA), Chuo-ku, Sagamihara City, Kanagawa 252-0222, Japan}
\address{${}^{56}$ Faculty of Law, Ryukoku University, Fushimi-ku, Kyoto City, Kyoto 612-8577, Japan}
\address{${}^{57}$ Department of Physics, University of Notre Dame, Notre Dame, IN 46556, USA}
\address{${}^{58}$ Graduate School of Science and Technology, Niigata University, Nishi-ku, Niigata City, Niigata 950-2181, Japan}
\address{${}^{59}$ Korea Astronomy and Space Science Institute (KASI), Yuseong-gu, Daejeon 34055, Korea}
\address{${}^{60}$ Department of Electronic Control Engineering, National Institute of Technology, Nagaoka College, Nagaoka City, Niigata 940-8532, Japan}
\address{${}^{61}$ Faculty of Science, Toho University, Funabashi City, Chiba 274-8510, Japan}
\address{${}^{62}$ Graduated School of Science and Technology/Institute for Quantum Studies, Chapman University Affiliated Scholar, Keio University, Kohoku-ku, Yokohama City, Kanagawa 223-8522, Japan}
\address{${}^{63}$ Accelerator Laboratory, High Energy Accelerator Research Organization (KEK), Tsukuba City, Ibaraki 305-0801, Japan}
\address{${}^{64}$ Faculty of Information Science and Technology, Osaka Institute of Technology, Hirakata City, Osaka 573-0196, Japan}
\address{${}^{65}$ Department of Space and Astronautical Science, The Graduate University for Advanced Studies (SOKENDAI), Sagamihara City, Kanagawa 252-5210, Japan}
\address{${}^{66}$ Research Center for Space Science, Advanced Research Laboratories, Tokyo City University, Setagaya, Tokyo 158-0082, Japan}
\address{${}^{67}$ Institute for Cosmic Ray Research (ICRR), Research Center for Cosmic Neutrinos (RCCN), The University of Tokyo, Kashiwa City, Chiba 277-8582, Japan}
\address{${}^{68}$ Department of Physics, Kyoto University, Sakyou-ku, Kyoto City, Kyoto 606-8502, Japan}
\address{${}^{69}$ Yukawa Institute for Theoretical Physics (YITP), Kyoto University, Sakyou-ku, Kyoto City, Kyoto 606-8502, Japan}
\address{${}^{70}$ Faculty of Science, Department of Physics, The Chinese University of Hong Kong, Shatin, N.T., Hong Kong}
\address{${}^{71}$ Applied Research Laboratory, High Energy Accelerator Research Organization (KEK), Tsukuba City, Ibaraki 305-0801, Japan}
\address{${}^{72}$ Department of Communications Engineering, National Defense Academy of Japan, Yokosuka City, Kanagawa 239-8686, Japan}
\address{${}^{73}$ Department of Physics, University of Florida, Gainesville, FL 32611, USA}
\address{${}^{74}$ Department of Information and Management  Systems Engineering, Nagaoka University of Technology, Nagaoka City, Niigata 940-2188, Japan}
\address{${}^{75}$ Department of Physics and Astronomy, Sejong University, Gwangjin-gu, Seoul 143-747, Korea}
\address{${}^{76}$ Department of Electrophysics, National Yang Ming Chiao Tung University, Hsinchu, Taiwan}
\address{${}^{77}$ Department of Physics, Rikkyo University, Toshima-ku, Tokyo 171-8501, Japan}


\ead{kjun0107@resceu.s.u-tokyo.ac.jp}

\begin{abstract}
In April 2020, KAGRA conducted its first science observation in combination with the GEO~600 detector (O3GK) for two weeks. 
According to the noise budget estimation, suspension control noise in the low frequency band and acoustic noise in the middle frequency band are identified as the dominant contribution. In this study, we show that such noise can be reduced in offline data analysis by utilizing a method called Independent Component Analysis (ICA). 
Here the ICA model is extended from the one studied in iKAGRA data analysis by incorporating frequency dependence while linearity and stationarity of the couplings are still assumed.
By using optimal witness sensors, 
those two dominant contributions are mitigated in the real observational data. 
We also analyze the stability of the transfer functions for whole two weeks data in order to investigate how the current subtraction method can be practically used in gravitational wave search.  
\end{abstract}

\section{Introduction}\label{sec:Introduction}
Since advanced LIGO (aLIGO) has achieved the first direct detection of gravitational wave (GW)~\cite{LIGOScientific:2016aoc} with their two detectors, 
more than 90 GW events~\cite{LIGOScientific:2018mvr,LIGOScientific:2020ibl,LIGOScientific:2021djp} have been detected by aLIGO and advanced Virgo. 
The GW detector KAGRA, constructed in Japan, has unique features compared to the aLIGO and advanced Virgo. KAGRA uses cryogenic mirrors to reduce thermal noise and it is located at the underground site to be isolated from seismic motions~\cite{KAGRA:2020tym}. KAGRA performed the first joint observation run (O3GK) with GEO600 in Germany, from April 7 to 21, 2020~\cite{LIGOScientific:2022myk,KAGRA:2022fgc} after the LIGO and Virgo O3 observation was suspended due to the COVID-19 pandemic. 
During O3GK, the median value of the binary-neutron-star (BNS) range of KAGRA, which is the observable distance of GWs from BNS coalescence, was about 0.6~Mpc \cite{LIGOScientific:2022myk}. 
While the continuous operation 
for two weeks itself has significant meaning for the KAGRA collaboration, one cannot expect astronomical findings from the KAGRA observation with its  current sensitivity. During the commissioning period before and after the O3GK, the contributions of various noise sources were investigated~\cite{KAGRA:2022fgc}, which provides useful information for noise reduction in KAGRA. 
Nevertheless, we should consider the possibility that noise partially remains in the data even after the noise reduction efforts in the instruments, which occasionally happens in experiments. There can be another possibility that a new noise source is found during the observation run which was buried in the previously dominant noise. 
In such situations, establishing a noise subtraction scheme in the data analysis after data acquisition\footnote{See also Ref.~\cite{Driggers:2011aa, DeRosa:2012bf, Meadors:2013lja} for the feed forward subtraction.} which, for example, makes use of the various witness sensors is highly beneficial for achieving the first detection of GW in KAGRA.  

Noise subtraction from GW strain data in offline analysis has been extensively studied, from the application of well-known filtering technique~\cite{Driggers:2011aa, Tiwari:2015ofa, LIGOScientific:2018kdd}, to the newly developed methods~\cite{Allen:1999wh, Davis:2018yrz, Vajente:2019ycy} and the machine learning~\cite{Ormiston:2020ele, Mogushi:2021deu, Yu:2021swq}. It is worthwhile mentioning that the strain sensitivity of LIGO Hanford during O2 was significantly improved by utilizing the Wiener filtering with the auxiliary sensors, {\it e.g.}, 
photodiodes that monitor beam motion and its size for beam jitter~\cite{LIGOScientific:2018kdd}. As for KAGRA data, application of Independent Component Analysis (ICA)~\cite{ICA1, ICA2, ICA3} was investigated with the initial KAGRA (iKAGRA) strain data and seismometers~\cite{KAGRA:2019cqm}. ICA is a method of signal processing which separates mixture of signals into statistically independent components by making use of non-Gaussianity of the sources. 
While the principle of ICA based on the statistical independence is so general that it is applicable to the non-linearly coupled system~\cite{Morisaki:2016sxs}, it makes the calculation more transparent compared to the machine learning.  
In the previous study~\cite{KAGRA:2019cqm}, we demonstrated that the seismic noise can be partially subtracted by the simplest model of ICA. Note that the model assuming instantaneous linear mixing, was implemented in two different ways and the optimal case turned out to be the time domain Wiener filtering~\cite{Wiener, Morisaki:2016sxs}. 

At the timing of iKAGRA, however, seismometers were the only environmental channels installed at the site. Since they are sensitive in the lower frequency band, our analysis cannot be applied to the $\mathcal{O}(100)$~Hz range, which is crucial to the GW observation. 
Since various environmental monitors including the ones sensitive to $\mathcal{O}(100)$~Hz range have been installed before O3GK~\cite{KAGRA:2020agh}, we can now apply ICA and investigate the noise subtraction in this frequency range.
Among those monitors, microphones are of our interest since the acoustic noise contamination has been reported~\cite{Washimi:2020slk}. 
In addition, the sensors in the local controls of the Vibration Isolation System (VIS) can be included in our analysis to subtract its control noise. In the iKAGRA data, the effect of the control noise was buried in the dominant seismic noise and was hardly observed due to the simplified suspension for the mirrors. With the current configuration of KAGRA, non-negligible contribution from the control noise has been estimated~\cite{KAGRA:2022fgc} and it is worth investigating whether it can be subtracted. With these consideration, in this work, we apply extended model of ICA with newly installed microphones and local sensors of VIS to the KAGRA O3GK data. 

As for the ICA model, convoluted linear mixing in time domain is considered to include the frequency dependence of the coupling.
The optimal subtraction in this case was shown to be the Wiener filtering for each mode~\cite{Morisaki:2016sxs} and here the discussion is generalized to multiple components. 
One can understand that the method estimates the so-called transfer function of the noise for each frequency. 
Among two possible ways of the estimation~\cite{Morisaki:2016sxs, Allen:1999wh}, we here utilize the one whose computational cost is lower and performance was investigated with the LIGO O2 data resulting in successful noise subtraction~\cite{Davis:2018yrz}. 
In order to ensure the validity of the method, we first 
apply it to the data in which acoustic noise was artificially injected and check whether the injected amount of noise is subtracted. 
Note that such a consistency check with hardware injection was not performed in the previous studies mentioned above. 
Then we analyze two prominent noise contributions mentioned above in the O3GK data and perform subtraction. 
We should note that, however, our model and formulation assumes the stationarity of the transfer function, which does not hold in general. In order to appropriately incorporate the current subtraction method into the data analysis flow, stability of the transfer function during the whole observation period needs to be quantified. 

The rest of paper is organized as follows. In Sec.~\ref{sec:Concept}, we make a brief review on the concept of ICA and show how the Wiener filtering in Fourier domain can be derived in our framework. In order to demonstrate validity of the subtraction formula, we perform the subtraction with the hardware injection data in Sec.~\ref{sec:injection}. Then we apply the method to the O3GK data and report the performance of the noise subtraction in Sec.~\ref{sec:O3GK}. 
In Sec.~\ref{TF_stability}, we show the stability of the estimated transfer function in specific frequency band for whole period of O3GK. 
Sec.~\ref{sec:conclusion} is devoted to the conclusion and the future prospects.


\section{Independent Component Analysis (ICA) with frequency dependence}\label{sec:Concept}
In this section, we first make a brief review on the concept of Independent Component Analysis (ICA)~\cite{ICA1,ICA2,ICA3} with the simplest mixing model. 
Then we reformulate the frequency dependent model discussed in Ref.~\cite{Morisaki:2016sxs} to the multiple components and consider its implementation to the data analysis. 

\subsection{Conceptual introduction of ICA and its application to noise subtraction}
To introduce the concept of ICA, we first consider the following simplest model where only two stationary sources $s_0(t)$ and $s_1(t)$ are involved.  
The observable $\vx(t)$ has the same number of components as the sources and can be written as 
\begin{eqnarray}
\vx(t)=\begin{pmatrix} x_0(t)\\
x_1(t)\end{pmatrix}=
\begin{pmatrix} a_{00}& a_{01}\\ a_{10} & a_{11}\end{pmatrix}
\begin{pmatrix} s_0(t)\\
s_1(t)\end{pmatrix}
=A\vs(t),  \label{vs}
\end{eqnarray}
where $A$ is here assumed to be a time independent matrix for simplicity. 
Generally speaking, the purpose of ICA is to recover $\vs(t)$ from $\vx(t)$ relying on the fact that $s_0(t)$ and $s_1(t)$ are statistically independent to each other and they obey non-Gaussian distributions\footnote{Precisely speaking, one of the sources is allowed to be a Gaussian variable.}.
For this linear mixing case, our goal is to find the inverse matrix of $A$ whose components are not known a priori. The estimation of the $A^{-1}$ is carried out in ICA as follows.
Let us consider linear transformation of the observable
\begin{eqnarray}
  \vy(t) 
  =\begin{pmatrix} y_0(t)\\
y_1(t)\end{pmatrix}=
\begin{pmatrix} w_{00}& w_{01}\\ w_{10} & w_{11}\end{pmatrix}
\begin{pmatrix} x_0(t)\\
x_1(t)\end{pmatrix}
  = W\vx(t),
\end{eqnarray}
which coincides with $\vs(t)$ if $W = A^{-1}$.
A probability
distribution function (PDF) of $\vy$, $p_y(\vy)$, 
is constructed from the observed PDF of $\vx$, $p_x(\vx)$, as 
\begin{eqnarray}
 p_y(\vy)\equiv ||W^{-1}||p_x(\vx),
\end{eqnarray}
where $||W^{-1}||$ denotes the determinant of $W^{-1}$.
Since the original sources are independent, 
$p_y(\vy)$ should coincide with the marginalized distribution function:
\begin{eqnarray}
 \bar{p}(\vy)\equiv \int p_y(y_0,y_1)dy_0\int p_y(y_0,y_1)dy_1=
\bar{p}_{0}(y_0)\bar{p}_{1}(y_1),
\end{eqnarray}
such that each component of the transformed variables $\vy$
is to be mutually independent in the statistical sense. 
In other words, $p_y(\vy) = \bar{p}(\vy)$ is understood as a necessary condition for $W = A^{-1}$.
Based on these arguments, the cost function of the estimation should be the one able to measure the ``distance" between the two arbitrary PDFs such as {\it Kullback-Leibler divergence}~\cite{KL}. Kullback-Leibler divergence is defined for two arbitrary
PDFs $p(\vy)$ and $q(\vy)$ as 
\begin{eqnarray}
 D[p(\vy); q(\vy)]=\int p(\vy)\ln \frac{p(\vy)}{q(\vy)}dy
=E_p\lkk \ln \frac{p(\vy)}{q(\vy)}\rkk, 
\end{eqnarray}
where $E_p[\cdot]$ denotes an expectation value with respect to a PDF $p$. In one implementation of ICA called FastICA~\cite{FastICA}, for example, $D[p(\vy), \bar{p}(\vy)]$ is approximately evaluated and minimized to find mutually independent variables. We should note that there can be a possibility that FastICA is trapped at local minima, which correspond to distributions mutually independent but different from the proper distribution of the sources.
To reach the correct answer, one may incorporate the information about the sources by explicitly assuming a concrete form of the PDF $\bar{q}(\vy)=q_1(y_1)q_2(y_2)$ and minimizing $D[p(\vy), \bar{q}(\vy)]$. 
When one considers noise subtraction from a main channel by using auxiliary monitors, however, 
the problem is significantly simplified due to the lack of the sensitivity of the monitors to the target signal in the main channel.

Let us identify the $x_0(t)$ as the output of a laser interferometer and $x_1(t)$ as the output of a witness channel. We can assume that $x_0(t)$ is composed of the GW signal $h(t)$ and stationary noise components $n(t)$, $k(t)$. Here $k(t)$ is the noise monitored by a witness channel $x_1(t)$ and $n(t)$ represents the noise components to which $x_1(t)$ is insensitive. Note that all the components are statistically independent. Then one can identify $s_1(t) = k(t)$ and $s_0(t) = h(t) + n(t)$, which results in the following form of the mixing matrix:
\begin{eqnarray}
A=\begin{pmatrix} a_{00}& a_{01}\\ 0 & a_{11}\end{pmatrix}. \label{aform}
\end{eqnarray}
Note that here $a_{01}$ can be rephrased as ``transfer function" of the noise $s_1$.
Since its inverse matrix also has a vanishing component $\left(A^{-1}\right)_{10} = 0$, we can focus on the specific form of the linear transformation $W$ as
\begin{eqnarray}
  W = \begin{pmatrix} w_{00}& w_{01}\\ 0 & w_{11}\end{pmatrix}, \label{wform}
\end{eqnarray}
to obtain statistically independent variables $\vy(t) = W\vx(t)$. 
By virtue of these assumptions~\eqref{aform} and \eqref{wform}, it turns out that we can easily construct such transformation only by considering two-point correlation of observable $\vx(t)$. It is straightforward to show that taking
\begin{eqnarray}
  w_{01}= -\frac{\la x_0x_1\ra}{\la x_1^2 \ra}w_{00}  \label{1211}
\end{eqnarray}
results in $\langle y_0y_1\rangle = 0$, 
where a bracket represents the ensemble average. Note that one can practically replace the ensemble average with temporal average owing to the assumed stationarity of the noise\footnote{Here we also assume $h \ll n, k$ so that the statistical property of the GW signal becomes irrelevant to the above discussion.}. 
What is special to the coupling~\eqref{aform} is that the condition $\langle y_0y_1\rangle = 0$ is necessary and sufficient to subtract $k(t)$ from $x_0(t)$, when $\langle s_0s_1\rangle = 0$ holds. This is a significant reduction of our problem since statistical independence generally involves higher order correlation function. For example, taking $y_0 = x_0 -\lmk\la x_0x_1\ra/\la x_1^2 \ra\rmk x_1$ and $y_1 = x_1$ yields $\langle y_0y_1\rangle = 0$. Indeed, 
\begin{eqnarray}
y_0(t) &=  x_0(t) -\frac{\la x_0x_1\ra}{\la x_1^2 \ra}x_1(t)\notag\\
&\simeq a_{00}(h(t)+n(t))+a_{01}k(t)-\frac{a_{01}a_{11}}{a_{11}^2}a_{11}k(t) = a_{00}(h(t)+n(t)),\label{Wiener}
\end{eqnarray}
does not contain $k(t)$ and thus does not correlate with $k(t)$ at any higher order, which satisfies the principle for the separation in ICA. Note that $a_{00}$ is irrelevant to our subtraction purpose.

One can see that Eq.~\eqref{Wiener} coincides with the Wiener filtering for two channels and it turns out to be a optimal subtraction method for the coupling~\eqref{aform} in the context of ICA.
In our previous work~\cite{KAGRA:2019cqm}, we showed that the above discussion is generalized to arbitrary number of witness channels $x_i(t)\ (i = 1, ..., n)$ and confirmed that the performance of this method corresponding to the Wiener filtering~\cite{Wiener} was better than FastICA in subtracting seismic noise from iKAGRA data. Based on this result, we extend the model of mixing matrix~\eqref{aform} by incorporating the effect of time delay, or the frequency dependence in the following.

\subsection{Frequency dependence and multiple components}
Next we extend the simplest model (\ref{aform}) to incorporate time delay as
\begin{eqnarray}
  \vx(t)= \int A(t,\tau)\vs(\tau)d\tau = \int A(t-\tau)\vs(\tau)d\tau , \label{vst}
\end{eqnarray}  
where $a_{12}(t-\tau) = 0$ and we have taken the matrix $A$ to be a function of time delay $t-\tau$
only assuming the stationarity.
Assuming that the delay time scale $\tau$ is sufficiently small compared to the temporal length $T$ of the data, the above coupling can be decomposed into the linear mixing of each Fourier mode~\cite{Morisaki:2016sxs}:
\begin{eqnarray}  
 \vtx(f) =
 \begin{pmatrix}
 \tx_0(f)\\ 
 \tx_1(f)
 \end{pmatrix}
 =
 \begin{pmatrix} 
 \ta_{00}(f) & \ta_{01}(f) \\ 
 0 & \ta_{11}(f)
 \end{pmatrix}
 \begin{pmatrix}
 \ts_0(f)\\ 
 \ts_1(f)
 \end{pmatrix}
 = \tilde{A}(f)\vts(f),\label{mix_f}
\end{eqnarray}
where a tilde represents the Fourier transformation of the quantities.
Note that such a frequency dependent linear coupling was actually observed in {\it e.g.} acoustic noise~\cite{Washimi:2020slk} and thus it is a reasonable extension to be explored. 
Now one can see that the mixing~\eqref{mix_f} has the same structure with the previous problem of the instantaneous linear mixing. 
Once again we wish to find statistically independent variables
\begin{eqnarray}  
  \vy(t)= \int W(t-\tau)\vx(\tau)d\tau,
\end{eqnarray}  
which is equivalent to
\begin{eqnarray}  
  \vty(f)=\tilde{W}(f)\vtx(f)= \begin{pmatrix} \tw_{00}(f) & \tw_{01}(f) \\ 0 & \tw_{11}(f)\end{pmatrix}\begin{pmatrix}
					    \tx_0(f)\\ \tx_1(f)\end{pmatrix}. \label{wformt}
\end{eqnarray}  
The independence of $y_0(t)$ and $y_1(t)$ requires that their correlation should vanish at an arbitrary order. However, as one can expect from the similar structure to the previous case, it is sufficient to eliminate the two-point correlation for the coupling model~\eqref{mix_f}:
\begin{eqnarray}  
    \langle y_0(t)y_1(t+\Delta t)\rangle
    &\propto
    \int df
    \langle\ty^{*}_0(f)\ty_1(f)\rangle
    e^{2\pi i f \Delta t}
    \notag\\
    &=0,\label{indep_memory}
\end{eqnarray}  
where $^*$ represents the complex conjugate and $\Delta t$ represents the time shift. 
Assuming that Eq.~\eqref{indep_memory} holds for any shift $\Delta t$, the condition for the statistical independence is satisfied when
\begin{eqnarray}  
\langle\ty_0^\ast(f)\ty_1(f)\rangle = 0\label{indep_freq}
\end{eqnarray}  
holds for each Fourier mode $f$.
In this way, the convoluted mixing~\eqref{vst} is now reduced to the problem for the stochastic variable $\ty_i(f)$ with the same condition to the previous one.
Because the normalization of $y_i$ is arbitrary, we can put $\tw_{00}(f)=\tw_{11}(f)=1$ and $\ty_1(f) = \tx_1(f)$. Then we find that the optimal subtraction formula turns out be the Wiener filtering for each frequency mode, namely, $\ty_0(f) = \tx_0(f) + \tw_{01}(f)\tx_1(f)$ with 
\begin{eqnarray}
  \tw_{01}(f)=-\frac{\langle \tx_0(f)\tx^*_1(f)\rangle}{\langle |\tx_1(f)|^2\rangle}
  \simeq
  -\frac{\ta_{01}(f)\ta_{11}^*(f)}{|\ta_{11}(f)|^2}.\label{TFfreq}
\end{eqnarray}
Note that the right hand side of Eq.~\eqref{TFfreq} is derived by using statistical independence $\langle\ts_0(f)\ts^*_1(f)\rangle = 0$ and results in $\ty_0(f) \simeq \ta_{00}\ts_0(f)$, which satisfies the principle of ICA.

As in the previous problem, the above discussion can be generalized to the case with multiple monitors $x_i(t)\ (i = 1, ..., n)$ where vanishing component of $\tilde{A}(f)$ then becomes $\ta_{i0}(f) = 0$. Consequently, $\langle \ty_0(f)\tx_i(f)\rangle = 0$ needs to be satisfied for mutual independence among $\vx(t)$. This is achieved through the Gram-Schmidt orthogonalization in terms of two point correlation between witness channels, as was done in Ref.~\cite{KAGRA:2019cqm} for the simplest instantaneous coupling. In this frequency dependent case, the cleaned strain data is given by
\begin{eqnarray}
\ty_0(f) = \tx_0(f)-\sum_{i=1}^n\frac{\langle \tx_0(f)\tx^{\prime*}_i(f)\rangle}{\langle |\tx^{\prime}_i(f)|^2\rangle}\tx^{\prime}_i(f),
\end{eqnarray}
where
\begin{eqnarray}
\tx_1^{\prime}(f) &= \tx_1(f),\\
\tx^{\prime}_i(f) &= \tx_i(f) - \sum_{j=1}^{i-1}\frac{\langle \tx_i(f)\tx^*_j(f)\rangle}{\langle |\tx_j(f)|^2\rangle}\tx^{\prime}_j(f).\label{diag}
\end{eqnarray}
This orthogonalization for the auxiliary monitors becomes important to avoid oversubtraction especially when the common noise is measured by different monitors. It is worthwhile mentioning that this Gram-Schmidt procedure can be free from the singularity problem which appears when one simultaneously treats all components of the covariance matrix~\cite{Allen:1999wh}.

Now we have obtained general expression for the frequency dependent subtraction, which was absent in Ref.~\cite{Morisaki:2016sxs}. While the above expression is written in the continuous limit, the real data sets are discretized and have finite duration and sampling rate.
In such a case, one can estimate the ensemble average of each Fourier mode in two different ways as explained below.

\subsection{Implementation of linear ICA with memory effect}\label{sec:implementation}
Here we summarize how the above method is practically implemented to the actual data analysis. The ensemble average of Fourier components can be evaluated in the following two ways.
As discussed in Ref.~\cite{Morisaki:2016sxs}, one may decompose the whole time series into $N_s$ smaller segments and obtain Fourier components $\tx_i(f;n_s)$ of each segment $n_s$\footnote{The length of the segments here is chosen to balance between the number of averaging and the frequency resolution.}. 
Then we can take average with respect to the segments $n_s$ to estimate the ensemble average of Fourier components. In other words, here the transfer function is estimated by the Welch's method. The subtraction can be performed for each chunk as
\begin{eqnarray}
\ty_0(f;n_s) = \tx_0(f;n_s)-\sum_{i=1}^n\frac{\langle \tx_0(f;n_s)\tx^{\prime*}_i(f;n_s)\rangle_{n_s}}{\langle |\tx^{\prime}_i(f;n_s)|^2\rangle_{n_s}}\tx^{\prime}_i(f;n_s),\label{Morisaki}
\end{eqnarray}
where $\langle\cdot\rangle_{n_s}$ represents the average with respect to the segments. 
We should note that the full length time series data can be properly reconstructed by allowing 50\% overlap when one decomposes the whole time series into the segments.
After the subtraction, each cleaned segment is multiplied by a Hann window in time domain and smoothly added together with 50\% overlap. In this way, we can provide the cleaned time series data with full length.

On the other hand, there is an alternative method to estimate $\tilde{W}(f)$, or the transfer function with a single Fourier transformation as described in~\cite{Allen:1999wh}. In this case, one splits the frequency space into bands with an adimensional width $F$ as
\begin{eqnarray}
f \in \left[f_b, f_{b+1}\right) \ \ {\rm with} \ \ f_b = \frac{bF}{T}.
\end{eqnarray}
Then the ensemble average can be replaced with taking average within each band $b$ and the subtraction formula becomes
\begin{eqnarray}
{\bm \ty}^{(b)}_0(f) = {\bm \tx}^{(b)}_0(f)-\sum_{i=1}^n\frac{\left( {\bm \tx}^{(b)}_0(f), {\bm \tx}^{\prime(b)*}_i(f)\right)}{\left( {\bm \tx}^{\prime(b)}_i(f), {\bm \tx}^{\prime(b)*}_i(f)\right)}
{\bm \tx}^{\prime(b)}_i(f),\label{BA}
\end{eqnarray}
where ${\bm \tx}_i^{(b)}(f) = \left(\tx_i(f_b), ..., \tx_i(f_{b+1}-1/T)\right)$ represents the vector of the Fourier modes within the frequency band $b$ and $\left(\cdot,\cdot\right)$ is the inner product of those vectors.
Since one can obtain the cleaned time series with a single inverse Fourier transformation in this case, this method can be implemented in much simpler way than the former one.
These two different implementation can be understood as the consequence of the duality of time and frequency and thus they are expected to have similar performance with each other.
In fact, Ref.~\cite{Davis:2018yrz} shows that this method can subtract comparable amount of noise with that subtracted by the  Wiener filtering~\cite{LIGOScientific:2018kdd} with lower computational cost.
Motivated with this study, here we adopt the latter method for the noise subtraction from the KAGRA O3GK data.

In this study, the analysis is based on gwpy~\cite{gwpy}.
Most of the FFTs are calculated with a Tukey window~\cite{Tukey} with $\alpha = 1/32$ in order to remove effects at the beginning and the end of the data stretch.
The band width $F$ is determined to balance between the frequency resolution and the sampling number for the average. As discussed in Ref.~\cite{Allen:1999wh}, the spurious correlation can be measured even for independent Gaussian variables, which decreases when $F$ becomes larger.
In order to discriminate the ``true" correlation from such artifacts, we need to set sufficiently large value of $F$, which was set to be at least $\mathcal{O}(10)$ in our analysis.
On the other hand, we also need to avoid over subtraction due to spurious correlation. This can be achieved by setting a threshold value for the coherence
\begin{eqnarray}
\tilde{\gamma}_{0i}(f) = \sqrt{\frac{\left|\left( {\bm \tx}^{(b)}_0(f), {\bm \tx}^{\prime(b)*}_i(f)\right)\right|^2}{\left( {\bm \tx}^{(b)}_0(f), {\bm \tx}^{(b)*}_0(f)\right)\left( {\bm \tx}^{\prime(b)}_i(f), {\bm \tx}^{\prime(b)*}_i(f)\right)}},\label{coherence}   
\end{eqnarray}
below which $\tw_{0i}(f)$, or the transfer function is set to be zero. Since $\langle\tilde{\gamma}^2\rangle = 1/F$ for independent random variables, we take threshold $N/F$ throughout this study, where $N$ is a constant of the order of unity. Note that we also apply this threshold to the diagonarization in Eq.~\eqref{diag}.

By comparing Eq.~\eqref{BA} and Eq.~\eqref{coherence}, one can see that the amount of the subtraction is characterized by the coherence.
In the following sections, we will show how this subtraction method works in the real KAGRA data by utilizing witness sensors which have large values of the coherence to the main channel of the detector.

\section{Testing the method: application of ICA to the acoustic injection data}\label{sec:injection}
Before moving to the analysis of the observational data, we here show the demonstration of noise subtraction by ICA with a single microphone. We use the data in which acoustic noise was injected by the speakers and check whether we can subtract the injected amount of the noise.
Since acoustic disturbance can easily be generated by using speakers and amplifiers, we can perform consistency check of the method with hardware injection which was not performed in the previous works including the iKAGRA data study~\cite{KAGRA:2019cqm}. 

The acoustic noise injection was performed on June 11, 2020, in the post-commissioning term of the O3GK. 
\begin{figure}[htbp] \centering
  \includegraphics[clip,width=13cm]{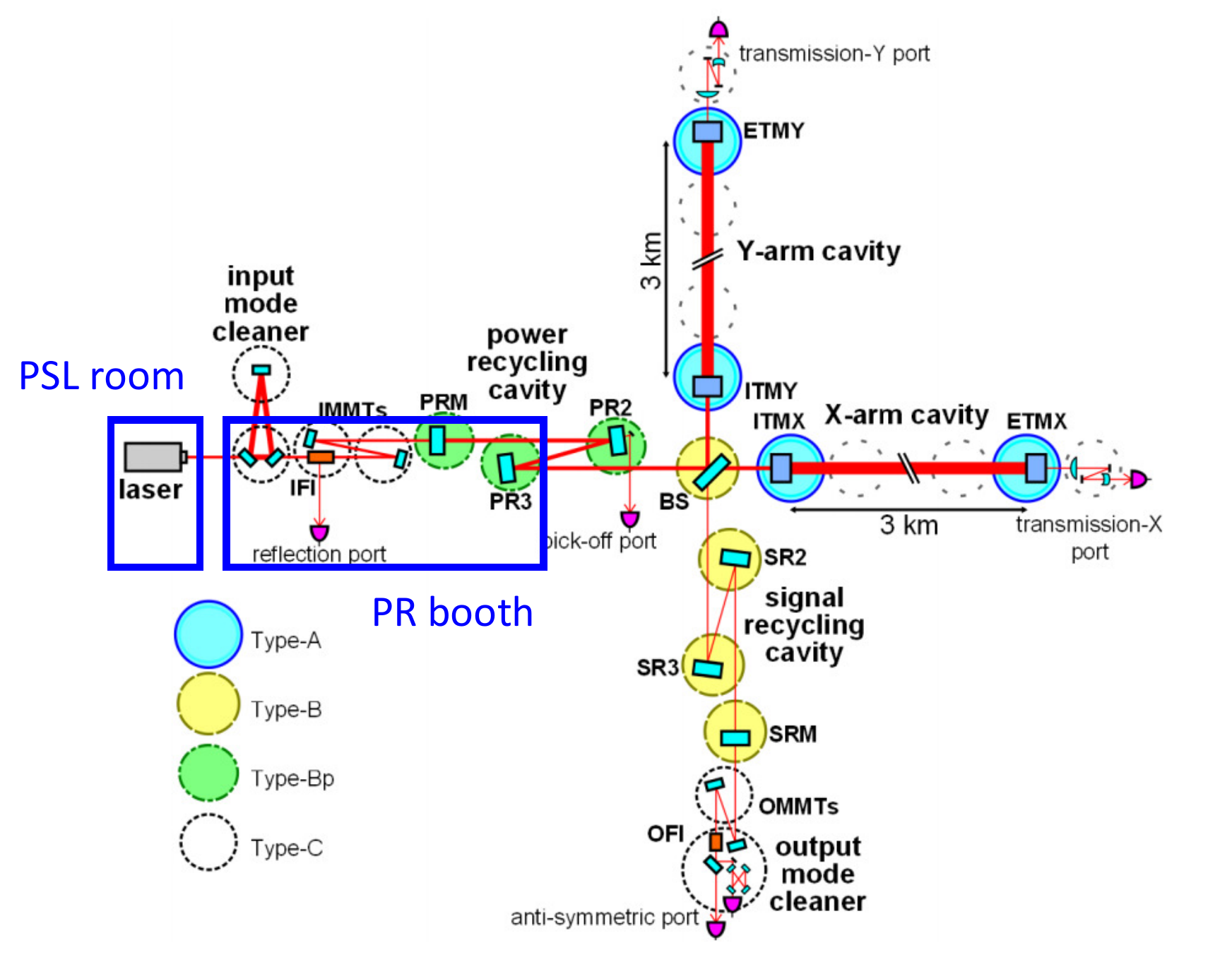}
  \caption{Schematic picture of the KAGRA configuration and the clean booth. The figure is taken from~\cite{KAGRA:2020tym}.}
  \label{fig:CleanBooth}
\end{figure} 
\Fref{fig:CleanBooth} is a schematic picture of KAGRA apparatus; 
main IR laser path, suspended mirrors, vacuum chambers, and clean booths.
According to the noise hunting in the pre-observation commissioning, it was found that the pre-mode cleaner in the PSL (pre-stabilized laser) room and 
the scattered light on the bellows between the IMC (input mode cleaner) and 
the IFI (input Faraday isolator) in the PR (power recycling) booth are sensitive to the acoustic disturbance.
Based on these observations, the acoustic noise was injected in the PSL room and the PR booth in order to investigate their couplings to the main channel, which is the differential arm length (DARM) strain channel in this case.
To quantify the frequency dependence and the (non-)linearity of the couplings, we injected single-frequency acoustic noise to both area with varying frequency from $70.0$~Hz to $1070.0$~Hz. For this wide frequency range, the DARM strain, shows mostly linear response to the acoustic injection in the PSL room, while some frequency conversion and sideband were observed for injection in the PR booth~\cite{Washimi:2020slk}. 
This result ensured that those data with the acoustic injection in the PSL room can be a suitable playground for testing the subtraction method for linearly coupled noise. 

\begin{figure}[htbp] 
\centering
  \includegraphics[clip,width=12cm]{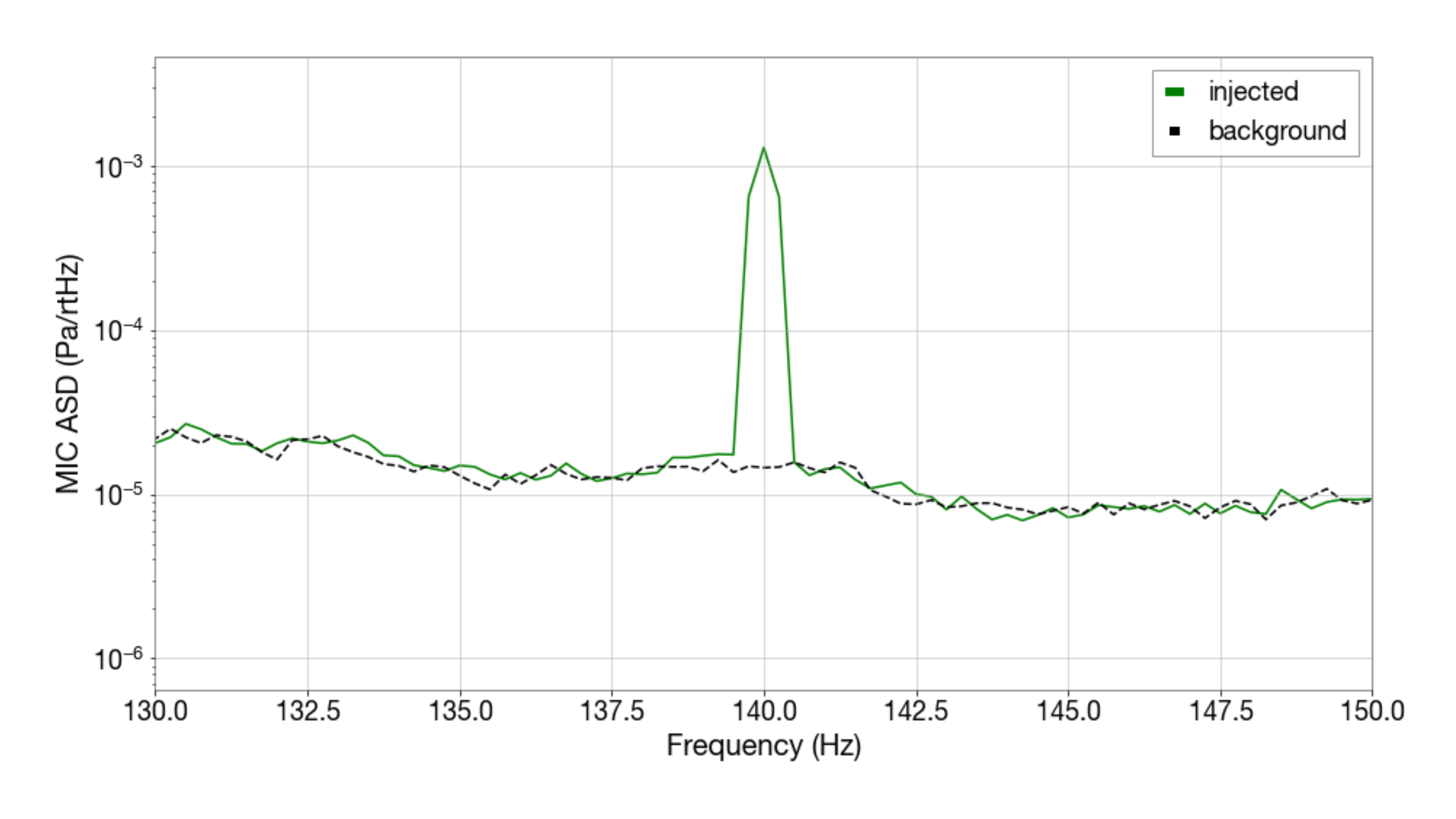}\\
  \includegraphics[clip,width=12cm]{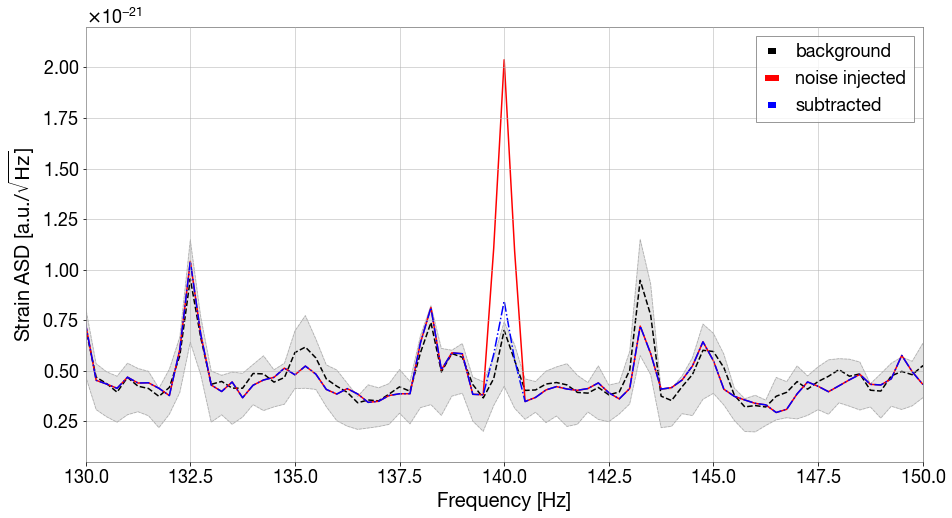}
  \caption{ASDs of the PSL microphone (top) and the DARM (bottom) strain during the single frequency acoustic noise injection in the PSL room. The gray-shaded region corresponds to the values within 25\% and 75\% percentiles. The threshold for ICA is set to $N = 3$ here, which results in the subtraction only at the injected frequency.}
    \label{fig:PSL_test}
\end{figure}

In order to check the validity of our method, we analyze the data with a fixed frequency noise injected in the PSL room for 128~seconds. 
As a witness sensor, we utilize a microphone installed at the table in the PSL room. 
Here the data with injection at 140~Hz is selected since the strain relatively well responds to the injection while there is no correlation between strain and the microphone without injection.
The amplitude spectral density (ASD) of the raw DARM strain and that of the DARM strain after processed with ICA are compared. 
Hereafter, each ASD is calculated using the Welch's method with a Hanning window and overlap = 50\%.
The FFT length and the bandwidth $F$ are chosen to 4 seconds and 32, respectively, for balancing between the frequency resolution and the variance of the estimated correlation.
The result is shown in Fig.~\ref{fig:PSL_test}. 
For a reference value, the background ASD of the DARM strain is also plotted, which is evaluated from a 2 minute data with no injection in the same lock segment. 
Note that the accurate calibration was not performed 
for these data sets and thus the ASD of the DARM strain is in an arbitrary unit. However, since their relative amplitude is preserved and it is sufficient for our purpose.
As one can see, the noise ASD at the injected frequency is reduced comparable to the background level after the linear subtraction. 
Here the accuracy in estimation of the transfer function is mainly limited by the small number of the samples in Eq.~\eqref{TFfreq} due to the short duration of the injection.
While the residual is slightly outside the 25\% and 75\% percentiles of the background ASD, 
it would be improved if the longer data were available. We thus conclude that the method can be consistently applied to the observational data.

\begin{figure}[htbp] 
\centering
  \includegraphics[clip,width=12cm]{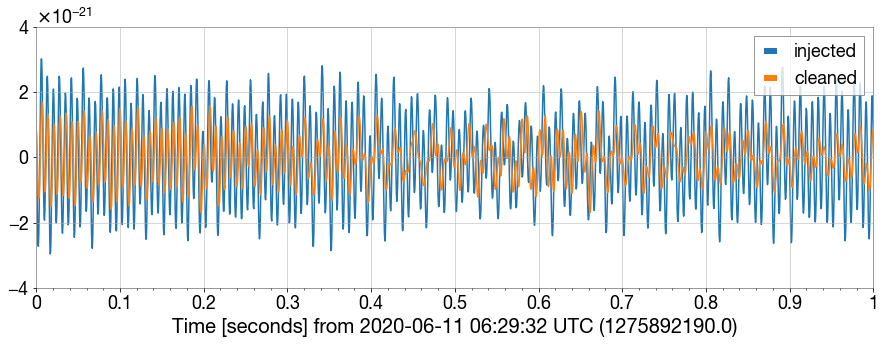}\\
  \caption{Time series of the DARM strain with bandpass $\pm$1~Hz around the injected frequency. After ICA, the amplitude corresponding to the 140~Hz component is reduced.}
    \label{fig:PSL_time}
\end{figure}

Note again that our subtraction method can provide the cleaned time series data, which is necessary for GW search pipelines. In Fig.~\ref{fig:PSL_time}, we plot time series of the noise injected strain data before and after the subtraction, both of which are bandpassed around the injected frequency as an illustration. While only 1 second is depicted in Fig.~\ref{fig:PSL_time}, full 2 minute cleaned data can be obtained after the subtraction\footnote{Both edge of the subtracted data is actually affected by the Tueky window for the one-shot FFT and one may dispose of such part. In the current parameter setup, however, it is only 4 seconds and more than 95\% of the original length is still available.}. One can see that the amplitude of the oscillation with corresponding frequency is attenuated.
Though we do not show its result here, the subtraction with the other implementation~\eqref{Morisaki} introduced in the former part of Sec.~\ref{sec:implementation} shows mostly similar performance to the above but with longer time for the calculation. For those who are interested, we summarize the result with the other implementation in \ref{app:morisaki}.
In consideration of the simplicity of the implementation and the computational cost, we intend to use only the method~\eqref{BA} in the following observational data analysis. 

\section{Noise subtraction from KAGRA O3GK data} \label{sec:O3GK}
\begin{figure}[htbp] 
\centering
  \includegraphics[clip,width=15cm]{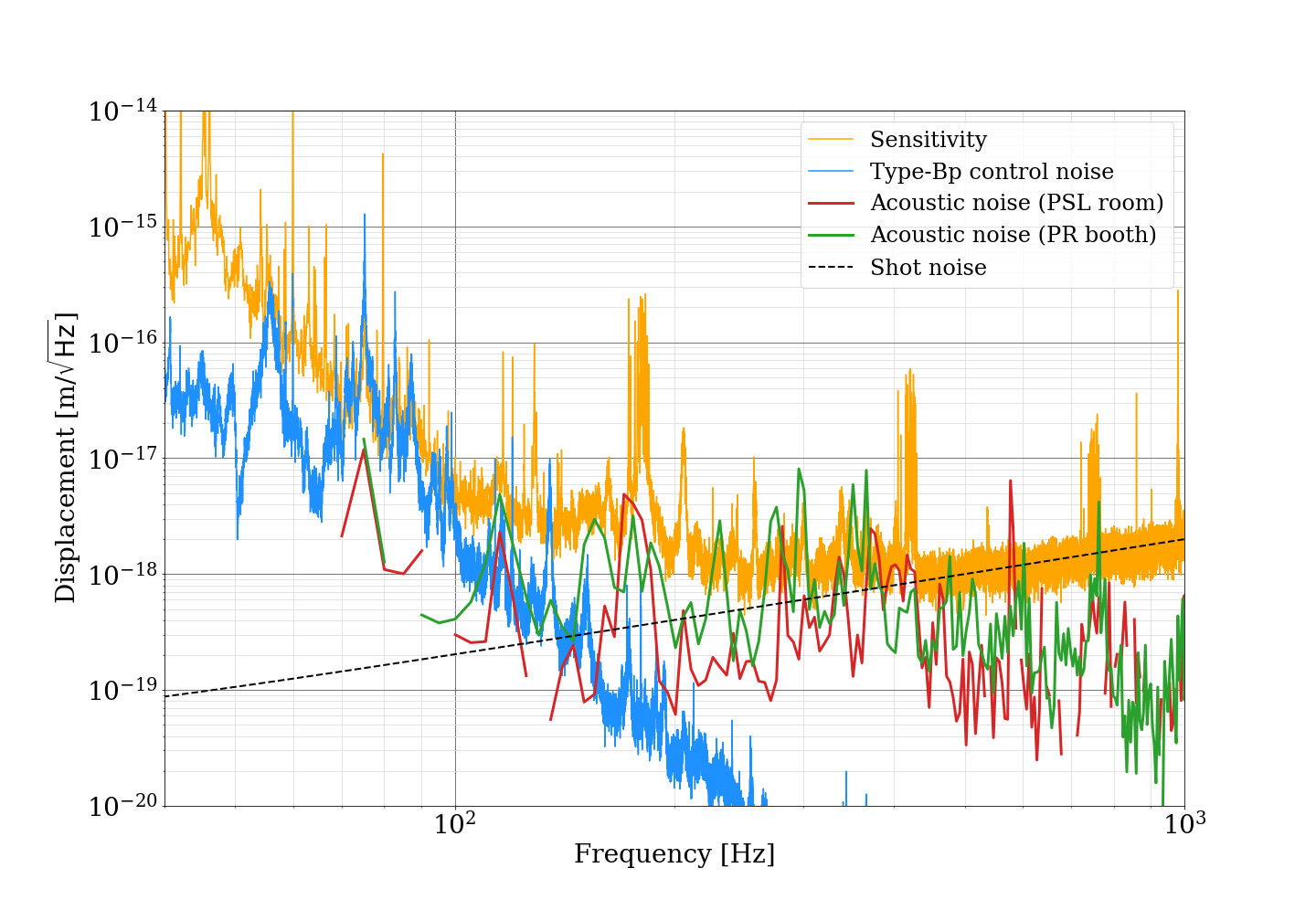}
  \caption{Typical amplitude spectral density (ASD) of the KAGRA strain sensitivity in the O3GK and the estimated contribution from the known noises~\cite{KAGRA:2022fgc,Washimi:2020slk}.
  Note that the acoustic noise in PR booth includes the effect of frequency conversion, which cannot be subtracted with the current linear method.}
    \label{fig:Sensitivity_O3GK}
\end{figure}

Now let us apply the subtraction method to the KAGRA's first observational data. As mentioned in the introduction, noise budget of the KAGRA O3GK data has been estimated. 
Among various sources of noise, here we focus on two specific noise contributions, namely, acoustic noise in the middle frequency range and suspension control noise (Type-Bp)
in the low frequency range, which are explained in the following. 
In Fig.~\ref{fig:Sensitivity_O3GK}, typical ASD of the DARM strain and estimated contribution from those noises are presented. As one can see, those estimated contributions becomes comparable to the typical ASD in the frequency range mentioned above.
We will show how those can be cleaned by ICA with the auxiliary monitors. 
In the following, we apply our method to 1024 seconds of the KAGRA O3GK data in the science mode~\cite{LIGOScientific:2022myk}. Here we set the parameters of ICA as the bandwidth $F = 256$ and the coherence threshold $N = 3$. In addition, we also limit the frequency range $(f_{\rm min}, f_{\rm max})$ in which the subtraction is performed depending on the auxiliary monitors. 

\subsection{Acoustic noise}\label{PEM}
As mentioned in the previous section, it was known that there was an unignorable coupling of the acoustic contamination in the PSL room and the PR booth to the main channel. 
While some parts of them were mitigated by reducing the corresponding sound sources and adding sound proofing~\cite{KAGRA:2020agh},
there are still remaining contributions~\cite{KAGRA:2022fgc}. It is worth investigating whether they can be subtracted by offline analysis.
The microphones installed at those places 
were known to have relatively large coherence to the DARM strain in some frequency bands and should be useful for ICA. With these motivation, here we utilize those two microphones simultaneously in our analysis. Based on the estimation in Fig.~\ref{fig:Sensitivity_O3GK}, we set $f_{\rm min} = 100$~Hz and $f_{\rm max} = 450$~Hz in this analysis.

\begin{figure}[htbp] \centering
\includegraphics[clip,width=14cm]{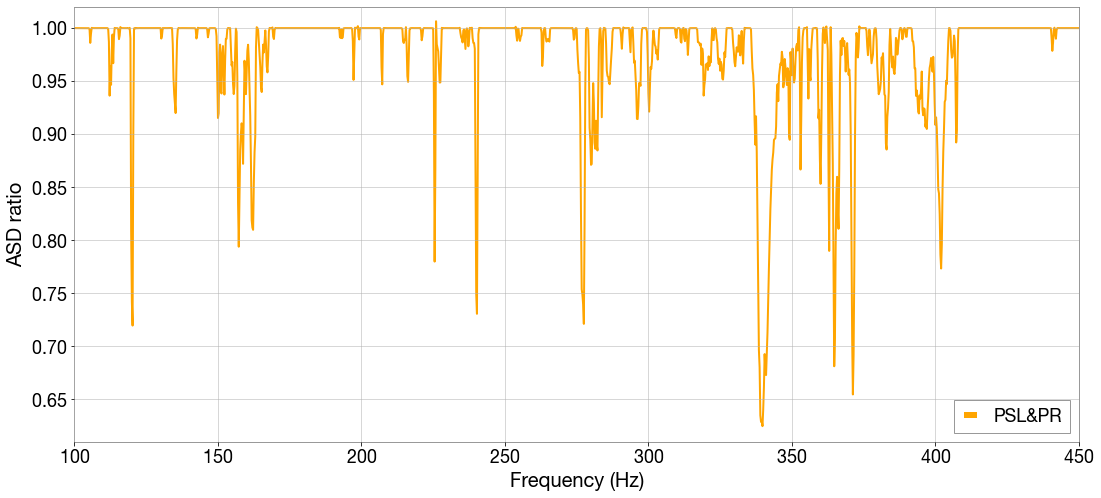}
  \caption{Ratio of the ASD of the DARM strain before and after ICA using microphones in PSL room and PR booth. 
  Noise reduction is most significant in the frequency range from 330~Hz to 405~Hz as expected from the noise budget~\cite{KAGRA:2022fgc}.}
    \label{fig:MIC_summary}
\end{figure}

\begin{figure}[htbp] \centering
\includegraphics[clip,width=14cm]{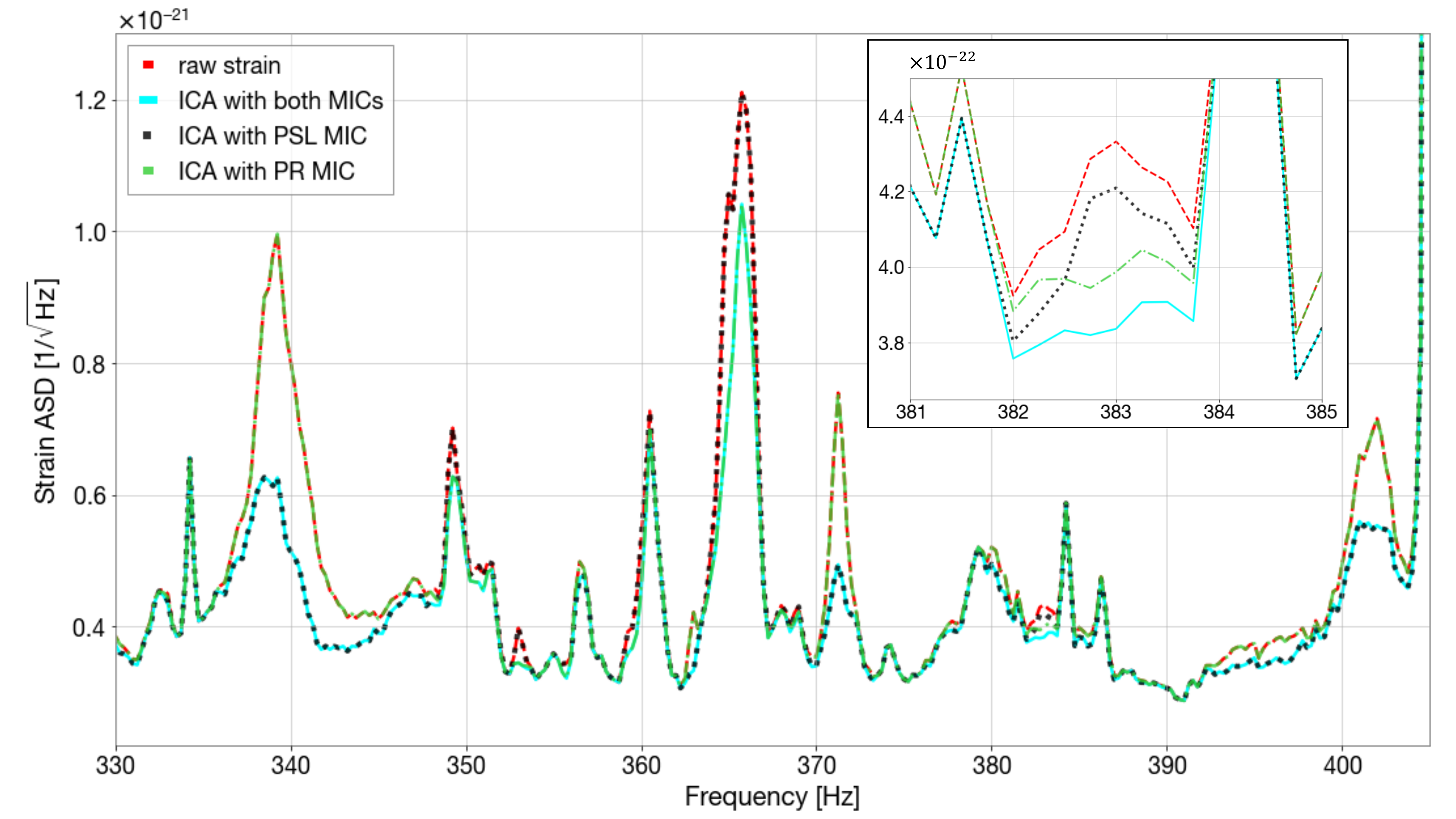}
  \caption{Comparison of ASD of the raw strain (red) and those of cleaned strain by ICA. 
  Black dotted line represents ICA with a microphone in PSL room while green line represents ICA with a microphone at PR booth. The cyan, ICA using both microphones, shows combined subtraction of each contribution.  
  }
    \label{fig:MIC_multi}
\end{figure}

In the same way as the demonstration in Sec.~\ref{sec:injection}, we compare the ASD of the raw strain and that of the strain processed with ICA.
In Fig.~\ref{fig:MIC_summary}, performance of ICA with those microphones is summarized. 
Here the ratio of the ASD before and after ICA is plotted and the smaller value means the larger amount of the subtraction. As one can see, the combination of those two channels results in broad reduction of noise.
Since the subtraction is effective in the frequency range where excess of the estimated noise contribution was observed, this result is consistent with the noise budget study.

Focusing on the frequency range from 330~Hz to 405~Hz where the subtraction seems significant, we show how the multiple channel analysis works compared to the single channel ICA. 
In Fig.~\ref{fig:MIC_multi}, the actual ASDs before and after ICA are shown. The results of ICA with a single microphone are simultaneously plotted for comparison.
It can be seen that the performance of single channel subtraction is not spoiled by the addition of another channel. For example, the contribution around 335-345~Hz and 370-373~Hz has linear correlation only with the PSL room microphone and thus subtracted by the single channel ICA with it. Inclusion of PR booth microphone, which does not linearly correlate with those contributions, does not affect the outcome of the multi-channel subtraction. On the other hand, the contribution around 383Hz seems to correlate with both microphones. Though it is small amount, combination of both channels results in further reduction in this case.
In summary, the outcome of multiple channel analysis is almost like a combination of the results of single microphone cases, which indicates that our implementation of the Gram-Schmidt orthogonalization~\eqref{diag} consistently works.
We would also like to emphasize that almost no noise is added through this subtraction process. We consider that this is owing both to the appropriate setting of a threshold and to the appropriate choice of the witness sensors for the analysis.
While there can be room for improvement in the implementation of, for example, threshold values, this result indicates that we should pay a careful attention to these microphones as monitors for subtractable noise in the forthcoming observation run.

\subsection{Type-Bp control noise}\label{sec:White}
Next, we consider the subtraction of suspension control noise by using local sensors.
The mirrors of the main interferometer are suspended with Vibration Isolation Systems (VISs) composed of multiple mechanical stages in order to isolate them from the seismic motion. On the other hand, the mechanical resonance would contaminate the detector around the resonant frequencies of the suspensions, which is designed to be lower than $\mathcal{O}(100)$~Hz. To suppress such vibration of the suspension, local damping controls were activated.
In KAGRA, there are four different types of VIS called Type-A, Type-B, Type-Bp and Type-C. Among them, noise from the local damping control in Type-Bp suspension turns out to be dominant over other suspension control noise between 50~Hz and 100~Hz~\cite{KAGRA:2022fgc}. 

As shown in Fig.~\ref{fig:CleanBooth}, the power-recycling mirrors (PRM, PR2 and PR3) are suspended by the Type-Bp suspension. 
Optical levers are used there as the local sensor monitoring the motion of the mirrors and they should be useful for the subtraction of control noise in the DARM strain. 
Therefore, we have utilized optical levers for those mirrors in this analysis.
For each mirrors, three degrees of freedom for the angular motion are monitored. Among them, sensors for pitch and yaw motion of the PRM and PR2 mirrors especially show larger coherence to the DARM strain and thus we select them for the noise subtraction. Here we set $f_{\rm min} = 40$~Hz and $f_{\rm max} = 100$~Hz for ICA.

\begin{figure}[htbp] \centering
    \includegraphics[clip,width=12cm]{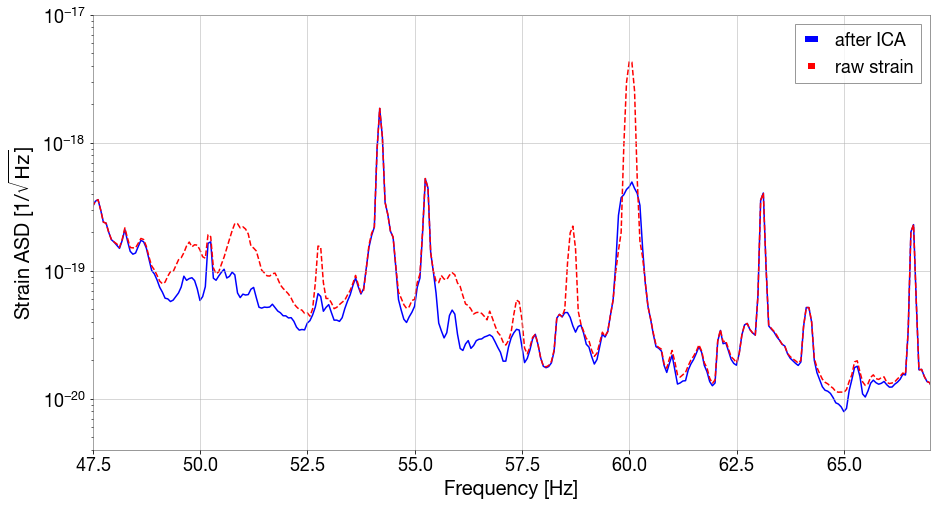}\\
    \includegraphics[clip,width=12cm]{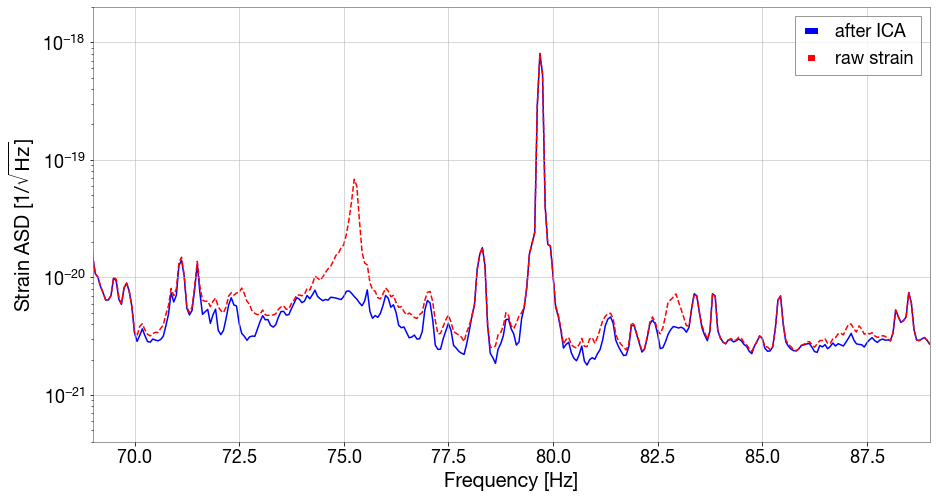}
  \caption{ASD of the raw strain (red) and that of the cleaned strain by ICA (blue) using the  optical levers related to the type-Bp suspension. Reduction of the control noise is mainly observed around 50~Hz(Top) and around 75~Hz(Bottom).}
    \label{fig:Oplevs}
\end{figure}

In Fig.~\ref{fig:Oplevs}, we compare ASD of the DARM strain before and after ICA with the Type-Bp optical levers. Here the FFT length is taken as 16~s.
Significant noise reduction is observed, for example, around 50~Hz and around 75~Hz\footnote{While optical levers are not the direct monitor of the power supply, the power line around 60~Hz are partially reduced due to the appearance of the same line in the optical levers outputs. For this line noise, one can perform subtraction with the optimal sensor for the power supply as shown in~\cite{Vajente:2019ycy}.}.
Compared with Fig.~\ref{fig:Sensitivity_O3GK}, again the subtraction seems consistent with the noise budget estimation which predicts large contribution in those frequency range. Similar to the above analysis with microphones, we have not observed contamination after the subtraction process.
Thus we can conclude that these sensors for the angular motion of the test masses are appropriate witness channels for the control noise subtraction.

Here we would like to briefly mention the possible origin of the noise around 50~Hz, which we found through this analysis. 
During the attempt to identify the witness channels for that contribution, we found that it can be subtracted also by the sensors installed at the tables in the PR booth, {\it e.g.} accelerometers. Since the type-Bp suspension is utilized for the mirrors in the PR booth, this result indicates that those sensors may observe the oscillation of the tables and its source also affects the control loops.
In this way, identification of the appropriate witness sensors may provide useful information of the noise sources.

\section{Stability of the transfer function during O3GK}\label{TF_stability}
Within a specific segment, we have demonstrated that some contributions of the noise in the KAGRA data during the O3GK run can be subtracted in the offline data analysis. However, our current method~\eqref{BA} assumes the stationarity of the system, or that of the transfer function~\eqref{TFfreq}, which does not hold in general for longer duration. 
While there were some attempts to take into account the non-stationarity, especially for slow variation~\cite{Davis:2018yrz, Vajente:2019ycy}, the problem generally becomes significantly complex.
Instead of seeking an improvement of the formalism itself by incorporating the non-stationarity, here we investigate how the current simple strategy can be appropriately applied to the whole period of the KAGRA O3GK data.
If the transfer function is approximately constant for some period, current subtraction method is practically applicable to that data segments. This motivates us to check how the estimated transfer function~\eqref{TFfreq} is actually stable during the whole O3GK period.
Note that here we check only its stability and do not consider subtraction procedure for whole two weeks data.
For this purpose, the transfer function can be evaluated simply by the Welch's method which has already been implemented in GWpy.
The level of environmental noise generally depends on time and the experimental site. Therefore, stability of the acoustic noise transfer function is worth being investigated.
Here we pick 57 data segments in which detector is continuously locked for more than 30 minutes. From them, we create 236 data chunks which have 30 minute duration and then evaluate the transfer function individually for each chunk with the FFT length = 32~s, overlap = 50\% and the Hanning window\footnote{Dependence on the window function is briefly discussed in \ref{app:morisaki} and the use of the Hanning window is shown to be appropriate for the estimation of the transfer function.}.

\begin{figure}[htbp] \centering
  \includegraphics[clip,width=15cm]{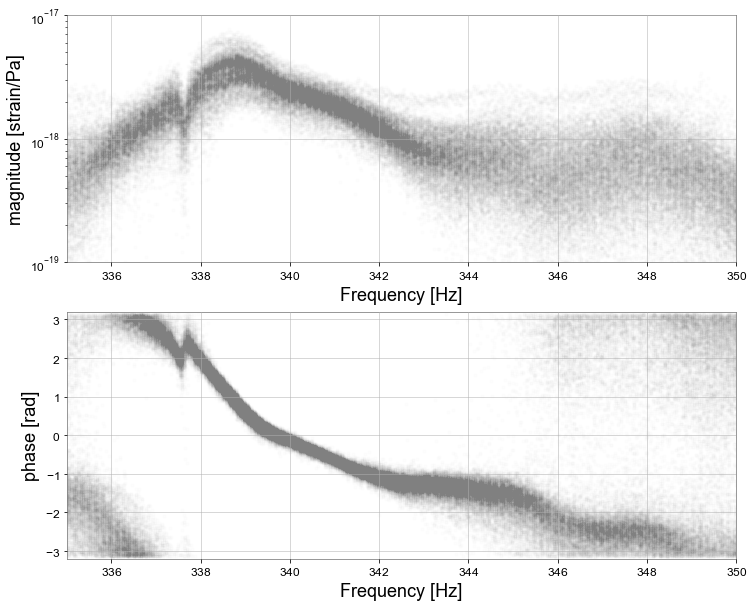}
  \caption{Scatter plot of the magnitude (Top panel) and the phase (Bottom panel) of the transfer function. Each black dot represents the evaluation within the 30 minute chunk.
  }
    \label{fig:TF_dist}
\end{figure}

As an example, the transfer function of the acoustic noise is individually evaluated for 236 data chunks by using the PSL room microphone.
Fig.~\ref{fig:TF_dist} shows the distribution 
of the transfer function for all 2 week data used in this analysis.
Note that the broad noise reduction took place around 340~Hz as depicted in Fig.~\ref{fig:MIC_summary}. 
One can clearly see the stable structure especially for the phase from 337~Hz to 345~Hz, while the magnitudes are also stable from 337~Hz to 342~Hz.
This behavior of the phase may indicate presence of localized sources of acoustic noise.
For such a stable structure, we may be able to subtract them in online analysis by using the transfer function measured, for example, in the first few minutes of the observation.

\begin{figure}[htbp] \centering
  \includegraphics[clip,width=12cm]{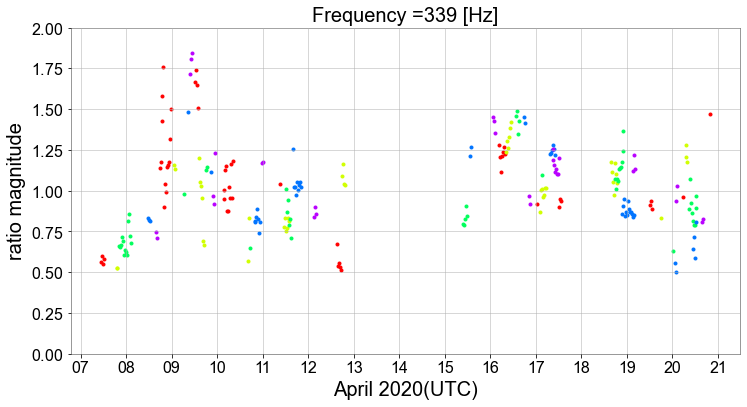}\\
  \includegraphics[clip,width=12cm]{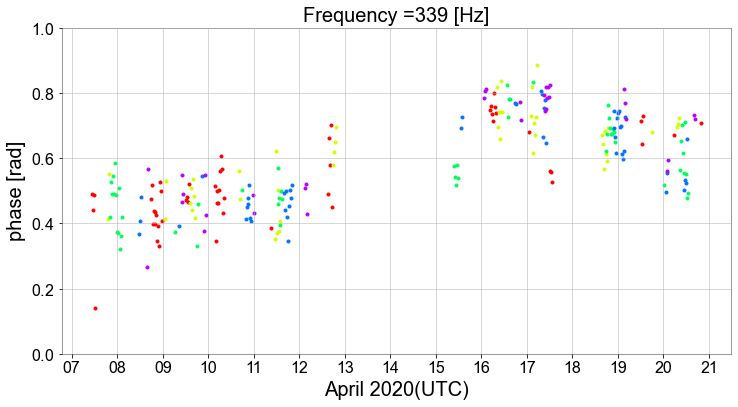}
  
  \caption{Time dependence of the transfer function of acoustic noise at 339~Hz. Here the magnitude is normalized by the median value. Each dot corresponds to the 30 minute chunk and the same color corresponds to the same lock segment.
  }
    \label{fig:TFcalender}
\end{figure}

For more quantitative discussion, the time evolution of this transfer function at a specific frequency bin is shown in Figures~\ref{fig:TFcalender}. Here we take $f = 339$~Hz as an example, which is within the frequency band where the transfer function seems stable. Here the same color corresponds to the same lock segment.
On one hand, one can see the aggregation of points of the similar color,
which indicates that the transfer function values within the same lock segment were stable in many cases. 
On the other hand, however, one can also see the case where magnitude of the transfer function largely changes in a relatively short time, for example, around the end of April 8th.
If the current method of subtraction is applied to such a lock segment at once, contamination is unavoidable due to incorrect estimation of transfer function. 
From this observation, we should conclude that time evolution of the transfer functions should be carefully checked even for the contribution which looks stable.

One may effectively take into account such a time evolution by separating the full-length data uniformly into, for example, 1024 second data and then smoothly connect the cleaned segments as in Ref.~\cite{Davis:2018yrz}. Our thorough analysis of the stability may indicate that, for the optimal subtraction, the length to divide needs to be varied depending on the segment. Such quantitative assessments can be done over time in offline analysis.

\section{Conclusion and future prospects}\label{sec:conclusion}
In this study, we performed linear noise subtraction from KAGRA O3GK data. Compared to the previous study where we applied two simplest models of ICA to the iKAGRA data~\cite{KAGRA:2019cqm}, we have extended the formulation of ICA by including frequency dependence in the noise coupling based on Ref.~\cite{Morisaki:2016sxs} and then generalized it into multi-channel case. Assuming the blindness of the witness channels to the other noise sources and the GW signal, the optimal subtraction corresponds to the Wiener filtering for each frequency mode. Note that the derivation is based on the statistical independence of the sources while the Wiener filter is usually derived to minimize the power of the residual. Such a difference should become evident when one removes the simplifying assumption and tries to derive more general expression of the ICA, for example, taking into account the non-linearlity~\cite{Morisaki:2016sxs}. 

As a consequence of the duality of the time and frequency, there are two possible implementations of the subtraction, or the estimation of the transfer function.
Since it can be implemented in a simpler way and reduced computational cost is achievable, we chose the one discussed in Ref.~\cite{Allen:1999wh} and applied to the LIGO O2 data~\cite{Davis:2018yrz} rather than the one discussed in Ref.~\cite{Morisaki:2016sxs}.
To show the applicability of extended ICA to the real KAGRA data, we first perform consistency check by using data with hardware noise injection. 
Linear excitation in the DARM strain due to the acoustic noise injection is successfully cleaned by ICA with a microphone and the residual is consistent with the data without noise injection. Therefore, we can safely apply our method to the KAGRA's observational data.

Based on the noise budget study~\cite{KAGRA:2022fgc} during the commissioning period, we focus on the two dominant noise contributions, namely, the acoustic noise and the suspension control noise.
As summarized in Sec.~\ref{sec:O3GK}, we have demonstrated how those contributions can be subtracted from KAGRA O3GK data by utilizing appropriate witness sensors. For acoustic noise, we identify two microphones which provides certain amount of the noise reduction. Multiple channel analysis proceeded with the Gram-Schmidt orthogonalization works well and we can properly combine contributions from those different microphones for the subtraction.
As for suspension control noise, we find that the local sensors in the control loop seem to be useful for the subtraction of the control noise. It is worth mentioning that we could get a glimpse of the potential origin of the type-Bp control noise from the PR booth microphones.
We should note that for both cases, result consistent with noise budget study is obtained and no spurious noise was added to the strain data through ICA.
For the detection of tiny GW signal, the latter point is significantly important, which can be achieved by finding the appropriate witness channels and avoiding the use of uncorrelated channels. Therefore, it would be better to select auxiliary monitors beforehand based on the coherence between them and the strain.

We also check the stability of the transfer function, which is a crucial assumption taken in our current ICA model. Focusing on the acoustic noise around 340~Hz related to the PSL room, as an example, the distribution and the time evolution of the transfer function is investigated for all 2 week data. While the structure of the transfer function seems relatively stable during whole observational period, there are a few lock segments in which the magnitude of the transfer function largely changes within a short period. To achieve the cleaning of the full-length data with current method, one should divide it into the smaller part in which the transfer function can be approximately regarded as constant. Or alternatively, the formulation of ICA incorporating the time dependence needs to be considered.
Such an issue will be further investigated in our future work.
Since current level of subtraction is insufficient to provide drastic change in SNR, effect on the signal detection is not investigated in this work.
After establishing method incorporating non-stationarity and also non-linearity, it is worthwhile to perform cleaning of the whole data segments and give a prediction for signal detection.

\section*{Acknowledgements}
This work was supported by MEXT, JSPS Leading-edge Research Infrastructure Program, JSPS Grant-in-Aid for Specially Promoted Research 26000005, JSPS Grant-in-Aid for Scientific Research on Innovative Areas 2905: JP17H06358, JP17H06361 and JP17H06364, JSPS Core-to-Core Program A. Advanced Research Networks, JSPS Grant-in-Aid for Scientific Research (S) 17H06133 and 20H05639 , JSPS Grant-in-Aid for Transformative Research Areas (A) 20A203: JP20H05854, the joint research program of the Institute for Cosmic Ray Research, University of Tokyo, National Research Foundation (NRF), Computing Infrastructure Project of KISTI-GSDC, Korea Astronomy and Space Science Institute (KASI), and Ministry of Science and ICT (MSIT) in Korea, Academia Sinica (AS), AS Grid Center (ASGC) and the Ministry of Science and Technology (MoST) in Taiwan under grants including AS-CDA-105-M06, Advanced Technology Center (ATC) of NAOJ, and Mechanical Engineering Center of KEK.

This work was partially supported by JSPS KAKENHI, Grant Number JP20J21866 (J.K.), 19J01299 and 20H05256 (T.W.). J.K.\ was supported by research program of the Leading Graduate Course for Frontiers of Mathematical Sciences and Physics (FMSP).

\appendix
\section{Alternative implementation}\label{app:morisaki}
Here we show how the alternative subtraction method~\eqref{Morisaki} works with the noise injected data discussed in Sec.~\ref{sec:injection}. 
As described in Sec.~\ref{sec:implementation}, this implementation estimates the transfer function with the Welch's method and thus there is a choice of the window function to be used in the analysis. 
\begin{figure}[htbp] 
\centering
  \includegraphics[clip,width=12cm]{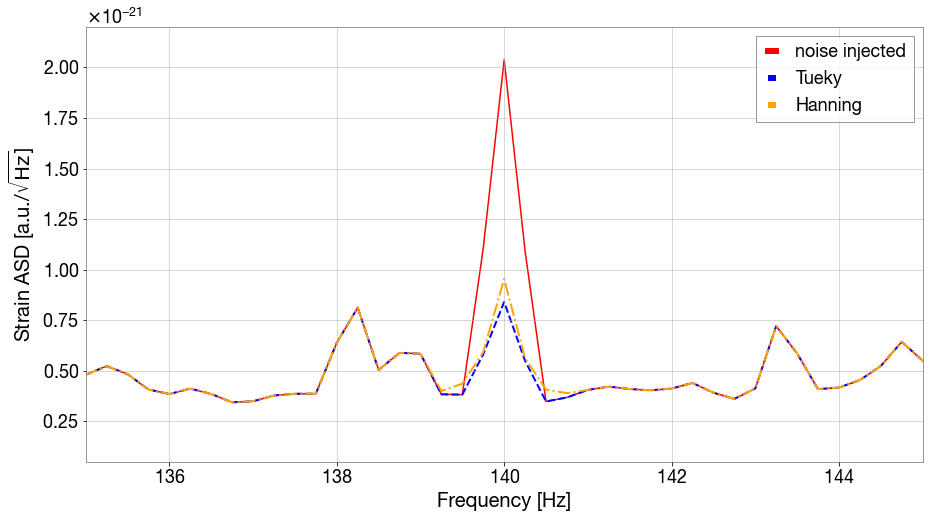}
  \caption{Subtraction of the injected noise with the alternative method~\eqref{Morisaki}. When the Tueky window is used, the result is hardly distinguished from the other implementation in Fig.~\ref{fig:PSL_test}. The residual becomes slightly larger for the Hanning window.}
    \label{fig:chunk_window}
\end{figure}

In Fig.~\ref{fig:chunk_window}, the result of subtraction~\eqref{Morisaki} is shown with two different window functions, which were multiplied to each chunk for FFT. 
Although the cleaned time series is not shown here, reconstruction of the full length is successfully done by the process discussed in Sec.~\ref{sec:implementation}.
Compared to the method in Fig.~\ref{fig:PSL_test}, the similar performance is obtained when the Tueky window is used. While there is a slight dependence on the choice of the window function, the peak height is reduced to the same level for the Hanning window. From this observation, our estimation of the transfer function in Sec.~\ref{TF_stability}, where the Hanning window is used for the Welch method, is validated.

We also compare the execution times when the two subtraction methods~\eqref{Morisaki} and~\eqref{BA} are applied to this 2 minute injection data. To get the cleaned 2 minute time series data, it takes 1.35s and 190ms respectively. As for the former method, FFT (and inverse FFT) needs to be performed chunk-by-chunk and the reconstruction of the full-length time series becomes complicated as described in Sec.~\ref{sec:implementation}. This time difference is expected to increase further as the data length increases and the data is divided into more chunks (while keeping the target frequency resolution fixed). 


\end{document}